# Concurrent Models for Function Execution

*Bob Diertens*

section Theory of Computer Science, Faculty of Science, University of Amsterdam

*ABSTRACT*

We derive an abstract computational model from a sequential computational model that is generally used for function execution. This abstract computational model allows for the concurrent execution of functions. We discuss concurrent models for function execution as implementations from the abstract computational model. We give an example of a particular concurrent function construct that can be implemented on a concurrent machine model using multi-threading. The result is a framework of computational models at different levels of abstraction that can be used in further development of concurrent computational models that deal with the problems inherent with concurrency.

## 1. Introduction

To execute a program written in a particular programming language, it is compiled into executable code for a particular machine. The machine is actually a machine model representing physical hardware, operation system, etc, or possibly a virtual machine. The compilation is done according to an execution model specific for the machine model. An execution model is an implementation of a computational model[1] which gives the essential rules for performing computations. The computational model must at least be adequate for expressing the operational semantics of the programming language. An overview of this all is given in Figure 1. For long the machine model was based on sequential execution of instructions. Programming languages were based on sequential execution of instructions as well, as were the computational models and the execution models.

With the introduction of support for concurrency in machine models, whether or not based on hardware capabilities, concurrency became available to be incorporated in programs. There are several forms in which support can be given for the use of concurrency. One of these forms is by using an add-on library that implements a set of primitives build on top of the concurrency capabilities provided by the machine model. A typical example of this is multi-threading [4]. A problem with the use of such libraries is that they may vary on different platform making the programs less portable. This problem can be solved by using a standard such as Posix [11], or Pthreads. A more important problem is that the programming language is still based on a computational model for sequential execution of instructions, and compilation is still based on a model for sequential execution of instructions. Compilers may generate efficient code that is correct for sequential execution, but incorrent for concurrent execution. This was already shown in [6] (1995) and again later in [5] (2005) and is caused by communications between threads through shared memory. The approach above results in the writing of a program, even if it clearly has independent parts that can be executed concurrently, based on a sequential execution model to which carefully concurrency is added, instead of onto a model that directly allows for concurrent execution transparently to the user. Furthermore, one has to keep control over the concurrent execution in order to avoid problems as race conditions, deadlocks, consistency of data, etc. So there is no transparency whatsoever concerning concurrency, making it very hard to write reliable software.

Another form to support concurrency is to extend existing programming languages with constructions that allow for some form of concurrency. These constructions are implemented on top of the existing computational and execution models using the same add-on libraries mentioned earlier. The problems with compilers generating code for sequential execution are hardly dealt with. Although this approach has some advantages over the use of add-on libraries, it is unclear how execution takes place and whether it can have different (unforeseen) results on different machine models.

---

1.    A computational model is also called an abstract machine model, although the terms can be considered different elsewhere.

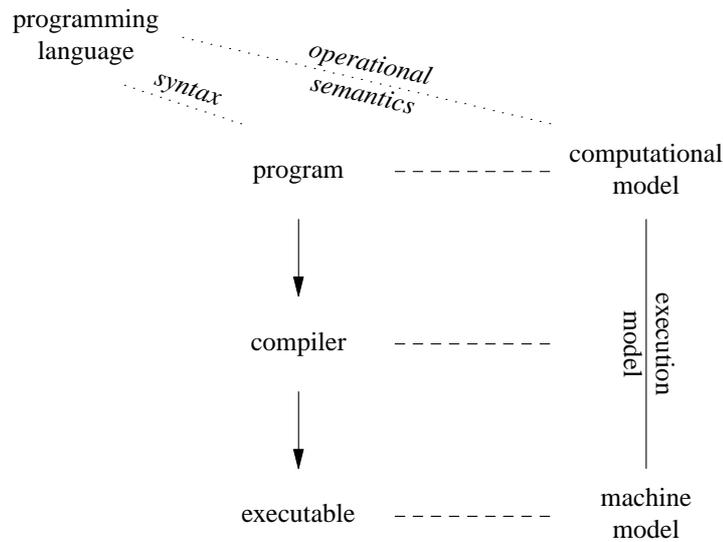

**Figure 1.** Overview of program execution related affairs

In both [6] and [5] it is stated that concurrency must be addressed at the language specification level and in compiler design. There are of course also new programming languages (or redesigned existing ones) which support concurrency that come with computational models and execution models which solve some of the problems, if not all, of concurrency, such as Java, C#, and Ada. But it is quite tricky to avoid problems with memory models for such a language as is shown in [16] and [15] for Java, and in [1] and [2] for a C++ standard.

In this article we focus on computational and execution models and leave out the problems with generating correct code for concurrent execution. We show that, just as the sequential execution model is one possible implementation of the sequential computational model, the latter just is one possible implementation of an abstract computational model that allows for concurrent execution of instructions. So instead of adding concurrency to the sequential execution model one should implement a concurrent execution model from the abstract computational model. Such an implementation should also keep concurrency as transparent as possible to the user in order to make writing concurrent software easier. We obtain an abstract computational model allowing concurrent execution of instructions from the sequential execution model via the sequential computational model by making the right abstractions. That abstract computational model can be used as base for exploring possible computational models allowing concurrent execution. From these computational models an implementation can be made for concurrent execution based on the capabilities of the machines.

Concurrency is applied at the instruction level in the form of concurrent execution of instructions in loop constructs. It is also applied at the function level, as is the case with multi-threading. Here, we focus on the concurrent execution of functions. A computational model for a programming language with functions should describe how these functions have to be executed. In this article we set out to develop computational models for programming languages with concurrent execution of functions.[2] We start with deriving a sequential computational model from a sequential execution model in section 2. In section 3 we abstract from the sequential computational model to obtain an abstract computational model for function execution with scheduling that allows for concurrent execution of functions. We explore a possible implementation for the abstract computational model still allowing concurrent execution of functions in section 4.

---

2. Often, the terms synchronous and asynchronous execution of functions are used instead of sequential and concurrent execution of functions.

## 2. Sequential Computational Model

A program written in a programming languages is translated to code for a particular machine in order to be executed. In general, this translation to machine code is based upon a model for sequential execution of instructions. We refer to program algebra [3] for information on instruction sequences. A sequence of instructions can be generalized through parameterization forming a function, making it possible to abstract from the implementation of the function. Calling a function from another instruction sequence is an essential element of most programming languages. Therefore, a model for the sequential execution of instructions must have a mechanism for implementing function calls. Here, we describe such a function call model up to a certain level of detail followed by a more generalized form of this model.

### 2.1 Sequential Function Execution Model

A function call can be described as a change of execution of instructions to the execution of instructions of the called function, and where the arguments of the function call are made available to the called function. After the execution of instructions of the called function is finished, the result is made available at the point of the function call, and the execution of instructions prior to the change is continued.

A function call is implemented in the machine model using some convention. We describe here a scheme that makes use of a stack. This scheme is based on a calling sequence for the C programming language as described in [12],[3] and the generated assembly code of some C compilers.

1. The arguments for the function to be called are pushed onto the stack.

2. The address of the instruction where execution has to be continued after the function call (return address) is pushed onto the stack.

3. Control is passed to the called function by setting the program counter to the start of the called function. Arguments of the function call are available through referencing in the stack.

   1. The contents of registers that are used inside the function are saved on the stack.

   2. Stack space is allocated for local variables of the function.

   3. The actual body of the function is executed.

   4. The return value is stored somewhere on the stack so that it can be obtained by the caller.

   5. Stack space is freed and register values are restored.

   6. The return address is popped of the stack.

   7. Control is given back to the caller of the function by setting the program counter to the return address.

4. The return value is taken from the stack.

5. All the values that had been pushed onto the stack are now popped from the stack (the stack is restored to its state before the function call) and execution of instructions continues.

The use of a stack in the above scheme for storing the values makes the recursive calls of functions possible. But it is not always necessary to use the stack for storing a particular value, for instance the return value can be saved in a dedicated register. The scheme shows that there is no special mechanism ivolved that takes care of the function call. Instead, the instructions for handling the function call are put inline with the other instructions.

---

3. Other conventions are possible too. More information can be found on http://en.wikipedia.org/wiki/Calling_convention and http://en.wikipedia.org/wiki/Call_stack.

## 2.2 Generic Model of Sequential Function Execution

In the scheme described above, the data on the stack is usually accessed through the register called *stack pointer*. The part of the stack that contains the data for a particular function call is called a *stack frame*, and holds the arguments for the function, the return address, and the local variables of the function. Alternatively, such a frame may be accessed through a special register called *frame pointer* pointing to the position of the frame on the stack, to allow for manipulation of the stack pointer during execution of the function. A stack frame consist actually of two parts, a part that is build up by the caller and a part that is build up by the function.

We can describe this more general without the use of a stack.

1. The arguments for the function and the return address are put in a frame.

2. The frame is saved in a place that is available to the function.

3. An environment for the function is set up.

4. Control is passed to the called function.

   1. The function builds up its own frame for storing local variables and saving contents of registers used inside the function.

   2. Arguments are taken from the frame.

   3. The actual body of the function is executed.

   4. The called function saves the return value in the frame build up by the caller.

   5. The function disposes its own frame.

   6. The return address is taken from the frame and control is given back to the caller of the function.

5. The environment of the function is taken down.

6. The return value is taken from the frame

7. The frame is disposed off.

Although we called it a stack frame, the actual use of a stack is not mentioned in the scheme above. The frame can be communicated to the function by pushing it onto the stack as a whole, but alternatives are also possible, such as communicating only the location of the frame through the stack or using a dedicated register for this.

## 2.3 Abstract Sequential Function Call

We use the above generic model for function execution to obtain an abstract model that hides the details of how a function call is implemented. From the caller's viewpoint in abstraction the call of the function can be seen as making the arguments available to the function. If we make a similar abstraction on the function side we get the following scheme.

1. Caller makes arguments and return address available.

2. An environment for the function is set up.

3. Control is passed to the function.

   1. Function initializes.

   2. Function gets arguments.

      3.   Function executes its body.

      4.   Function makes return value available.

      5.   Function cleans up.

      6.   Control is given back.

   4.   Environment is taken down.

   5.   Caller gets return value.

The abstract model given here is the sequential computational model we are looking for. It abstracts from all possible details of implementation and focusses on the sequential computation of function calls.

## 3.  Abstract Computational Model for Function Execution

The sequential computational model for function execution that we obtained in the previous section describes a model can be seen as one possible sequential implementation of a concurrent computational model. In this section we describe how a concurrent computational model can be obtained from the sequential computational model by abstracting from the details of how a function is scheduled for execution.

### 3.1  Abstract Function Call

In abstraction, from the caller's viewpoint the call of the function can be seen as the sending of arguments to the function and the receiving of the return value. If we make a similar abstraction on the function side we get the following scheme.

   1.   Caller sends arguments to function.

      1.   Function receives arguments.

      2.   Function executes its body.

      3.   Function sends return value to caller.

   2.   Caller receives return value and continue execution.

The scheme above does not allow for recursive function calls. It also assumes there is a function present for receiving the arguments and return address, but the function first has to come into existence. In order to solve this a mechanism is needed that creates an instance of the function when it is called, corresponding to the setup of an environment for the function in the sequential computational model.

### 3.2  Function Instance Controller

In the previous section we recognized the need for a mechanism that creates instances of functions. Such a mechanism also has to take care of the return values from these instances and the destruction of the instances after execution. Below we show a scheme in which this mechanism is called the controller.

   1.   Caller sends function-name and arguments to controller.

      1.   Controller receives function-name and arguments.

      2.   Controller creates an instance of the function.

      3.   Controller sends arguments to function.

         1.   Function receives arguments.

         2.   Function executes its body.

         3.   Function sends return value to controller.

      4.   Controller receives return value.

      5.   Controller disposes the instance of the function.

      6.   Controller sends return value to caller.

   2.   Caller receives return value and continues execution.

The abstract viewpoint for the caller and the function are still there, but now the controller acts as intermediate. It is also possible that on instantiation a function is supplied with the arguments, making the sending of arguments by the controller and the receiving of them by the function redundant. The scheme presented here is more general from the viewpoint of the function, in that it is not necessarilly the controller supplying the function with arguments. With this scheme it is also possible to consider the sending and receiving of the arguments as part of the instantiation of the function.

We obtained a scheme in which the controller operates in concurrency with sequential instruction execution. This gives the possibility to separate the concerns of the communication between the two and their internal operation. Abstracting from details of the internal working makes the model more comprehensive and at the same time it results in a model that allows for other implementations as well.

### 3.3  Scheduling of Functions

In the abstract model for function execution we use a controller that acts as intermediate for the communication with the instances of functions. In this section, we relax some constraints in this model to obtain a model in which the role of the controller becomes the scheduling of functions.

We can relax the constraints that a function has to wait for the value returned by the called function and that the controller immediately has to take action on a call of a function. Execution of instructions might as well continue until a certain point where the return value is needed. The controller may wait with creating an instance of a function until some criterion is met, such as the availability of resources, the moment the return value is needed, etc. This is typically the concept of a scheduler.

The tasks of a scheduler consists of receiving function call requests, controlling function execution, and sending return values back to the callers of the functions. Controlling function execution consists of the following steps.

   1.   Scheduler sets up an environment and makes the arguments available in this environment.

   2.   The function is executed in the environment.

   3.   The environment is taken down.

In this scheme there may be more than one function waiting for execution and they are not necessarilly executed in the order in which they are called. To deliver the right return values it is necessary to identify the calls and the instances. This can be done by providing instances of functions with identifiers which they have to send along with their messages.

### 3.4  Concurrent Execution

With the scheme presented above concurrent execution of function instances is possible. Since it is not necessary to wait for the return value of a function, other instructions can be executed concurrently with the execution of the function. At the same time other function calls can be executed as well, resulting in massive concurrent execution of functions. For parallel execution to happen, more than one sequential instruction executor has to be available. The scheduler has to use these sequential instruction executors as resources for execution of functions.

## 4.  Concurrent Models for Function Execution

We can implement a concurrent computational model from the abstract computational model that resembles the operational semantics of a programming language supporting a particular form of concurrent execution of functions. From the concurrent computational model a concurrent execution model can be implemented for a certain concurrent machine model, as depicted in Figure 2. The figure gives just an abstract view.

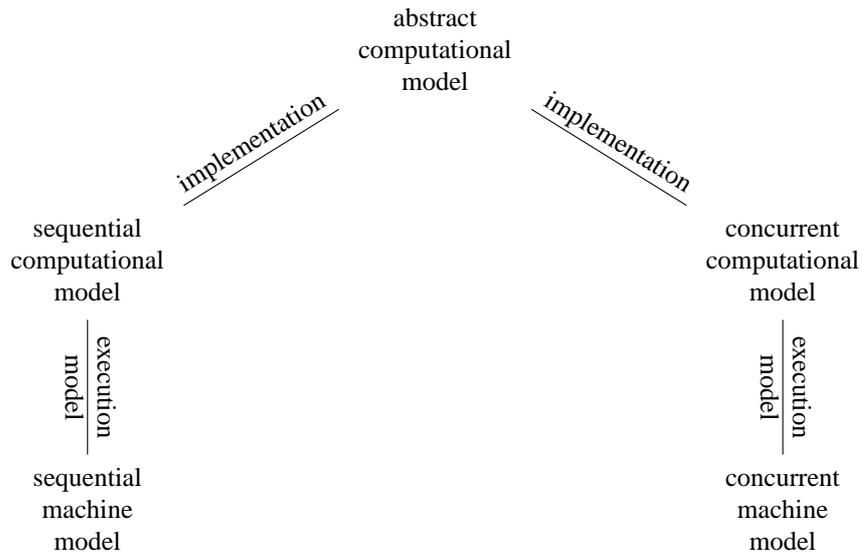

**Figure 2.** Framework of computational models

There are of course more concurrent computational models possible as implementation from the abstract computational model, each with one or more concurrent execution models for one or more concurrent machine models. This is also the case for sequential computational models. Furthermore, there are more levels of abstraction possible for the computational model as well as for the machine model.

In the framework described above we deal with the execution of functions on different levels of abstraction. Each lower level is a refinement of the level above, until reaching the target machine model. In the concurrent computational model we also have to deal with the problems caused by the concurrent execution of functions, such as problems with shared memory. With the sequential computational model this is not necessary as these problems dissappear due the inline scheduling of functions.

In the following section we briefly sketch the concurrent models for a particular concurrent function execution construct in a programming language.

## 4.1 Concurrent Function Execution

We can implement concurrent functions with constructs that separate the invocation of the function from the receiving of the return value from the function. In between the invocation of the function and the receiving of the return value the caller can do something else in concurrency with the execution of the function. A programming language may support this with the following construction.

```
fid = invoke(f(...))
      .
      .
      .
r = wait(fid)
```

Here, `invoke` returns an identifier which can be used by `wait` to wait for the function with this identifier to finish after which it returns the return value of the function. A sequential form of this can then be written as `r = wait(invoke(f(...)))`, or even shortened to `r = f(...)`.

In a computational model both the `invoke()` and `wait()` can be considered separate requests to the scheduler. On an `invoke()` the scheduler reacts with sending back an identifier and taking care of the execution of the function. On a `wait()` the scheduler checks if the function with the given identifier has finished. If it has, the scheduler sends the return value back. Otherwise the scheduler has to take care of sending the return value back after the function finishes. In the mean time the caller waits for the return value.

A program containing these constructs can be compiled to executable code for a particular machine model that supports concurrent execution, provided that there exists an execution model for it. If a machine model supports multi-threading, execution of concurrent functions can be mapped onto existing libraries of functions implementing threads. Sequential function calls can still be implemented using an execution model for sequential execution of functions.

The result is that we lifted the functionality of multi-threading libraries to the level of the programming languages and pushed the implementation of the libraries to the machine model. This increases transparency and makes code more portable across different machines. It is also possible for concurrent functions to be implemented using different forms of concurrency provided by machine models. The advantage of this approach is that we can deal with other extensions separately on a lower level of abstraction. For instance, we can further refine the computational model by adding communication between functions in the form of message passing or through shared memory.

## 5. Related Work

In many cases concurrency is added to existing programming languages by providing add-on libraries or by adding constructs supporting concurrency. The computational model in these cases however is still based on a sequential computational model. Even entirely new concurrent programming languages are often based on a sequential computational model. In some cases a clear distinction is made between synchronous and asynchronous execution of functions, what amounts to sequential and concurrent execution of functions. Although this distinction is made, there is hardly any mentioning of a computational model dealing with concurrency. It mostly comes down to building on top of a machine model that is extended with the functionality of an add-on library. So, concurrency is dealt with on a too low level of abstraction. The framework we presented in this article is in contrast with the work mentioned above, as it deals with concurrency on a higher level of abstraction. We have not seen any other work that tries to model concurrency at an higher level of abstraction.

We have not found any mentioning of using different scheduling mechanism for the execution of functions. Although different scheduling mechanisms are used, they are not characterized as such. Because of this, there is no proper understanding of the impact of using different scheduling mechanisms and of their interaction when used intermixed. Our framework can be used to study the use of different scheduling mechanism and the consequences they may have on the execution of functions.

There are many different machine architecture supporting parallel execution. Supporting a particular concurrent programming language on a wide range of these architectures is impractible. At the same time, many programs are customized to achieve higher performance on a particular architecture, making them less portable. A solution to this is a machine model[4] that can be implemented on many different architectures. Examples of such machine models are the Parallel Random Access Machine (PRAM) model [9], the Bulk-Synchronous Parallel (BSP) model [17], and the LogP model [7]. But there are many others and variants. Comparisons and analysis of some models can be found in [10], [8], [14], and [13]. The work on machine models fits perfectly on our framework, since it allows us to focus on concurrency on a higher level of abstraction instead of on details of specific machines. Furthermore, it increases both portability and scalability of the programs using concurrency.

## 6. Conclusions

In this article we derived an abstract computational model for the execution of functions. We started with the traditional sequential execution model for function execution from which we obtained a sequential computational model by abstracting from the details of function call implementation. By further abstraction of how a function is scheduled for execution, we obtained an abstract computational model that allows for the concurrent execution of functions.

---

4.    Such machine models are often called parallel computation models, but in this context we prefer to call them (abstract) machine models.

We have shown that with abstraction and relaxing constraints a model for execution of functions can be obtained in which function scheduling plays a key role. This model has as a possible implementation inline scheduling, the original stack-based function execution model we started with. Moreover, this model allows for concurrent execution of instructions, and therefore it is suitable as model for implementation of concurrent software.

Furthermore, we discussed concurrent models for function execution as implementations from the abstract computational model and we gave an example of a particular concurrent function construct in a programming language that could be implemented on a concurrent machine model using multi-threading. This showed that we can lift the functionality of multi-threading libraries to the level of the programming languages and with that push the implementation of the libraries to the machine model.

The overall result of computational models at different levels of abstraction can be used as a framework for further development of concurrent computational models that deal with the problems inherent with concurrency. The main advantage of this is that we can properly handle these problems on the right level of abstraction. The framework shows that it should not be decided what parts of a sequential system can be done concurrently, but what parts of a concurrent system can or should be done sequentially.